\documentclass[sigconf, nonacm]{acmart}
\usepackage{xspace}
\usepackage[ruled,vlined]{algorithm2e}
\usepackage{booktabs}   % for \toprule, \midrule, \bottomrule, \cmidrule
\usepackage{graphicx}   % for \resizebox
\usepackage{multirow}
\usepackage{array}
\usepackage{arydshln}

\newcommand\vldbdoi{XX.XX/XXX.XX}

\newcommand\vldbavailabilityurl{https://github.com/uiuc-kang-lab/sodium}
\newcommand\paradigmname{\textsc{Sodium}\xspace}
\newcommand\datasetname{\textsc{SodiumBench}\xspace}
\newcommand\sysname{\textsc{SodiumAgent}\xspace}
\newcommand\algname{\textsc{ATP-BFS}\xspace}
\newcommand{\minihead}[1]{{\vspace{.5em}\noindent\textbf{#1.}}}
\begin{document}
\title{\paradigmname: From Open Web Data to Queryable Databases}

%%
%% The "author" command and its associated commands are used to define the authors and their affiliations.
\author{Chuxuan Hu}
\affiliation{%
  \institution{UIUC}
  % \city{Urbana}
  % \state{IL}
}
\email{chuxuan3@illinois.edu}

\author{Philip Li}
\affiliation{%
  \institution{UIUC}
  % \city{Urbana}
  % \state{IL}
}
\email{philipl2@illinois.edu}

\author{Maxwell Yang}
\affiliation{%
  \institution{UIUC}
  % \city{Urbana}
  % \state{IL}
}
\email{my37@illinois.edu}

\author{Daniel Kang}
\affiliation{%
  \institution{UIUC}
  % \city{Urbana}
  % \state{IL}
}
\email{ddkang@illinois.edu}

\begin{abstract}
During research, domain experts often ask analytical questions whose answers require integrating data from a wide range of web sources. Thus, they must spend substantial effort searching, extracting, and organizing raw data before analysis can begin. We formalize this process as the \paradigmname task, where we conceptualize open domains such as the web as latent databases that must be systematically instantiated to support downstream querying.
Solving \paradigmname requires \emph{(1)} conducting in-depth and specialized exploration of the open web, which is further strengthened by \emph{(2)} exploiting structural correlations for systematic information extraction and \emph{(3)} integrating collected information into coherent, queryable database instances.

To quantify the challenges in automating \paradigmname, we construct \datasetname, a benchmark of 
105 tasks derived from published academic papers across 6 domains, where systems are tasked with exploring the open web to collect and aggregate data from diverse sources into structured tables.
Existing systems struggle with \paradigmname tasks: we evaluate 6 advanced AI agents on \datasetname, with the strongest baseline achieving only 46.5\% accuracy. 
To bridge this gap, we develop \sysname, a multi-agent system composed of a web explorer and a cache manager. 
Powered by our proposed \algname algorithm and optimized through principled management of cached sources and navigation paths, \sysname conducts deep and comprehensive web exploration and performs structurally coherent information extraction.
\sysname achieves 91.1\% accuracy on \datasetname, outperforming the strongest baseline by approximately $2\times$ and the weakest by up to $73\times$.
\end{abstract}

\maketitle

% \pagestyle{\vldbpagestyle}
% \begingroup\small\noindent\raggedright\textbf{PVLDB Reference Format:}\\
% \vldbauthors. \vldbtitle. PVLDB, \vldbvolume(\vldbissue): \vldbpages, \vldbyear.\\
% \href{https://doi.org/\vldbdoi}{doi:\vldbdoi}
% \endgroup
% \begingroup
% \renewcommand\thefootnote{}\footnote{\noindent
% This work is licensed under the Creative Commons BY-NC-ND 4.0 International License. Visit \url{https://creativecommons.org/licenses/by-nc-nd/4.0/} to view a copy of this license. For any use beyond those covered by this license, obtain permission by emailing \href{mailto:info@vldb.org}{info@vldb.org}. Copyright is held by the owner/author(s). Publication rights licensed to the VLDB Endowment. \\
% \raggedright Proceedings of the VLDB Endowment, Vol. \vldbvolume, No. \vldbissue\ %
% ISSN 2150-8097. \\
% \href{https://doi.org/\vldbdoi}{doi:\vldbdoi} \\
% }\addtocounter{footnote}{-1}\endgroup

\ifdefempty{\vldbavailabilityurl}{}{
\vspace{.3cm}
\begingroup\small\noindent\raggedright\textbf{PVLDB Artifact Availability:}\\
The source code, data, and/or other artifacts have been made available at \url{\vldbavailabilityurl}.
\endgroup
}

\section{Introduction}
\label{sec:intro}
Domain experts decompose their research objectives into smaller analytical subproblems. At each stage, they formulate a concrete analytical question together with a provisional database schema. However, the corresponding values are often scattered across deeply nested webpages within specialized websites.
For example, demographic research frequently investigates questions whose underlying data are publicly available on government statistical portals \cite{Berghammer_51_34, repec:dem:demres:v:51:y:2024:i:37,Treleaven_51_32, Tsui_50_45, Moretti_50_42, Raz-Yurovich_50_34, Kan_50_21, Muniz_50_17, Vermeulen_49_20, Mogi_49_16}. To study ``disability-free grandparenthood in Italy between 1998 and 2016'' ($Q_1$)~\cite{Moretti_50_42}, a researcher defines a table with primary key \texttt{year} (1998-2016) and columns representing household metrics. Yet the required values are distributed across multiple datasets and webpages. 
Before any analysis can proceed,
the researcher needs to navigate the Italian National Institute of Statistics (Istat) website~\cite{istat_esploradati} to access data from the Family and Social Subjects (FSS) surveys of different years, as well as the corresponding Italian life tables, and manually aggregate them into a unified table.
As a result, substantial overhead effort is spent on searching, extracting, and organizing raw values prior to performing the intended analytical task \cite{8440815,Hazzan2023,data-science-life-cycle}.
Given the rapid progress of large language models (LLMs) and AI agents on complex reasoning and interaction tasks \cite{ag2_docs,autogpt_repo,microsoft_autogen,openai2024researchbot,google_deep_research_2024}, automating this process becomes both feasible and compelling.

\begin{figure}[t]
    \graphicspath{{figures/}}
    \centering
    \includegraphics[width=\columnwidth]{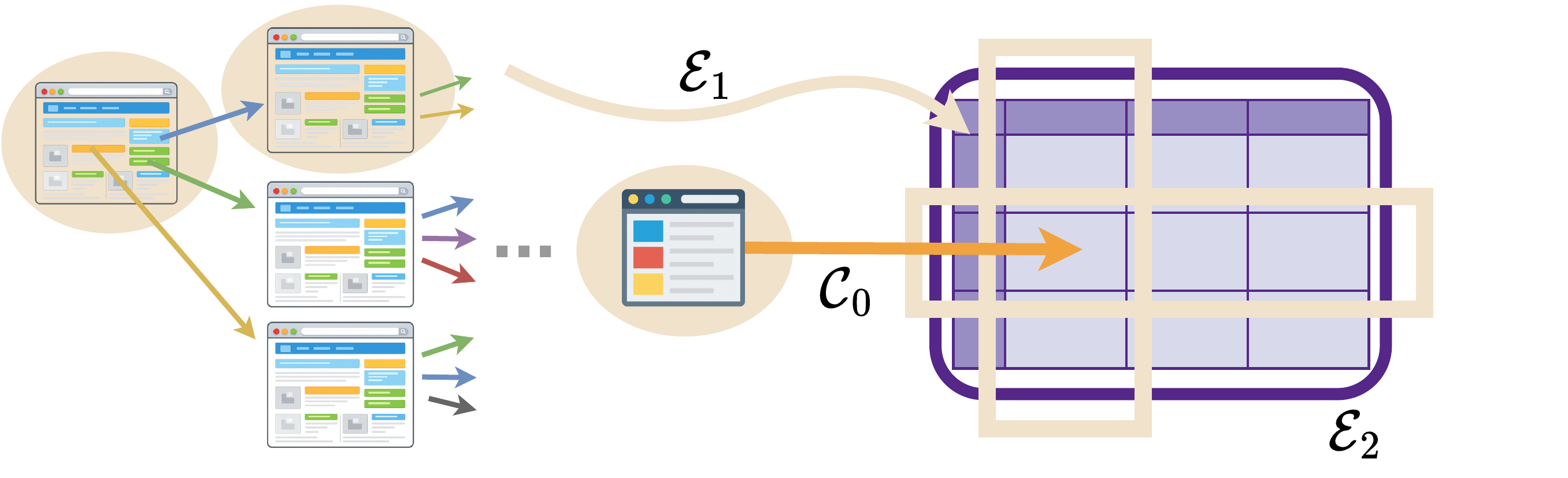}
\caption{
% Visualization of the three core dimensions in solving a \paradigmname task:
% ($\mathcal{C}$0) identifies relevant values through multi-step exploration of heterogeneous webpages,
% ($\mathcal{E}$1) exploits similarities across related entries to infer reusable navigation patterns, detect mismatched sources, and enforce cross-cell consistency, and ($\mathcal{E}$2) integrates the collected values into a coherent database instance aligned with the target schema, enabling downstream analytical querying.
Solving \paradigmname is driven by ($\mathcal{C}$0) in-depth exploration of specialized websites, and strengthened through ($\mathcal{E}$1) exploiting structural correlations for systematic information extraction and ($\mathcal{E}$2) integrating collected information into coherent, queryable database instances.
}
    \label{fig:capabilities}%
\end{figure}

We formalize this process as the \paradigmname task, which focuses on \textbf{S}tructuring \textbf{O}pen \textbf{D}oma\textbf{I}n \textbf{U}nstructured Data into \textbf{M}aterialized Databases, in Section~\ref{sec:usecase}.
In \paradigmname, users pose a concrete analytical query over a particular organization’s or platform’s official website and specify the expected database schema. The system is required to populate this given schema with values discovered from the open web.
\paradigmname models open domains such as the web as latent databases and formalizes the process of systematically materializing a database instance.
As we show in Figure \ref{fig:capabilities}, solving a \paradigmname task relies on a fundamental open-domain search capability ($\mathcal{C}_0$), complemented by two structure-aware extensions that enable cross-cell reasoning and coherent database materialization ($\mathcal{E}_1$, $\mathcal{E}_2$):

\minihead{($\mathcal{C}$0) In-depth information discovery}
A system must go beyond surface-level page inspection and conduct hierarchical, multi-step navigation to locate fine-grained data embedded within complex web portals. For example, to retrieve the household data needed for $Q_1$, one must first locate the Istat data exploration portal, from which it selects the ``Population and Households'' category among 20 available options, then navigate to the ``Households'' subcategory among 10 subcategories, and finally identify the correct ``Type of Households'' dataset among 22 datasets.

\minihead{($\mathcal{E}$1) Identification and utilization of structural correlations across cells}
% A system must recognize dependencies and relationships among table cells and leverage this structural information during retrieval and reasoning. 
Webpages often follow consistent organizational patterns, where relationships across entities (columns) and primary keys (rows) can be mapped to regular navigation patterns across webpages. Thus, structural dependencies across table cells can guide the filling of related entries by aligning web structures with table structures.
For example, after retrieving a value under the ``Type of Households'' dataset in $Q_1$, locating values under the ``Household Size'' dataset does not require restarting the search process; instead, the system can directly navigate to the corresponding ``Households'' subcategory page by reusing the previously identified organizational structure (\emph{column dependency}).
Similarly, once values for the year 1998 are retrieved, the remaining years are located in structurally similar webpages (\emph{row dependency}).
Exploiting these structural regularities optimizes exploration and supports scalable, efficient, and systematic information extraction.
% after identifying a plausible navigation path for one portion of $Q_1$, the system should examine whether other cells with similar semantic roles follow the same structural pattern. If related values are expected to reside in parallel datasets but instead appear under a different branch, this discrepancy may indicate that the previously selected source is suboptimal or semantically mismatched. In such cases, structural reasoning allows the system to revise earlier choices and redirect exploration toward a more consistent path.
% Thus, structural correlations serve two critical purposes: they help identify the optimal navigation strategy aligned with the table schema, and they enable cross-cell consistency checks that improve reliability. By jointly reasoning over structurally dependent entries, the system can detect inconsistencies, correct potential errors, and ensure that collected values originate from structurally coherent and semantically comparable sources.

\minihead{($\mathcal{E}$2) Organization of collected information into structured, queryable databases}
The populated table serves as input for downstream analysis and must therefore maintain internal consistency. For example, in $Q_1$, the datasets on Istat provide two types of measures (``Thousands value'' and ``Per 100 households with the same characteristics''), and the system must ensure consistent formatting and representation across these measures.

\vspace{.5em}

To systematically evaluate \paradigmname, we construct \datasetname, a benchmark of 105 analytical queries together with their target table schema and specialized domains derived from 48 published academic papers spanning 6 domains (Section~\ref{sec:data}). 
The scale and diversity of \datasetname underscore that structuring open-domain web data into materialized databases is a widespread requirement in practical research workflows. 
In each \datasetname task, systems are tasked with exploring the open web in the specialized domain to populate the table based on the query.

We evaluate 6 state-of-the-art agents with the most advanced web search capabilities on \datasetname in Section \ref{sec:exp}. Our evaluation shows that they perform poorly on \paradigmname tasks, with AG2~\cite{ag2_docs} achieving the highest accuracy at 46.5\%, and Open Deep Research~\cite{opendeepresearch} achieving the lowest at 1.2\%. 
These results empirically demonstrate that existing systems do not align with the core demands of \paradigmname. In fact, none of the existing systems fully realizes any of $\mathcal{C}_0$, $\mathcal{E}_1$, or $\mathcal{E}_2$, as we elaborate through the following detailed analysis of the systems with the closest related functionality.

First, while existing web search engines~\cite{fastapi_docs}, LLM-integrated web search tools~\cite{openai_web_search_docs,claude_web_search_docs,gemini_web_search_docs}, and web agents~\cite{he2024webvoyager} have shown promising performance~\cite{jin2025searchr1trainingllmsreason}, they focus on general-purpose and surface-level exploration of the open web. In contrast, $\mathcal{C}$0 requires comprehensive, in-depth explorations of specialized web domains. With the exception of WebVoyager, which attempts to enhance exploration but ultimately traverses only a narrow subset of available navigation paths and achieves less than 3\% accuracy on \datasetname, existing systems and agent frameworks~\cite{ag2_docs,autogpt_repo,microsoft_autogen} do not optimize at the search tool level, which fundamentally limits their ability to perform deep, domain-aware exploration.

Second, existing information retrieval (IR) processes, including the retrieval phase of RAG-based systems~\cite{gao2024retrievalaugmentedgenerationlargelanguage} and even table-population workflows~\cite{lakefill}, treat each retrieval independently, ignoring structural dependencies across related searches. Although the underlying data pool exhibits internal structure (e.g., graph~\cite{NEURIPS2024_efaf1c97}), these systems fail to link related retrievals by exploiting shared structural regularities, as embodied in $\mathcal{E}$1.

Third, existing LLM-based data management systems, whether operating over open-domain~\cite{dsdd,Hu_2025} or local data~\cite{lai2025kramabenchbenchmarkaisystems, Hu_2024}, assume that the input data is already organized and well structured. Thus, they bypass the central challenge addressed by $\mathcal{E}_2$: guaranteeing the coherence and consistency during materializing databases.

% Existing web search tools and agents fall short in two key aspects required for open-domain analytics:
% \emph{(1) information collection} and \emph{(2) information organization}.
% First, existing systems do not explore the web deeply enough to mine the comprehensive and interrelated information necessary to construct a database.
% Second, they do not take the schema-driven nature of databases into account when constructing database instances.
% Specifically, existing open-domain data collection systems typically assume that web data is already pre-structured, enabling users to directly download or extract tables~\cite{Hu_2025}.
% Moreover, information retrieval and textual lookup systems \cite{he2024webvoyager} primarily focus on retrieving relevant documents or passages, without accounting for the structural requirements of queryable databases.
% As a result, these systems fail to support analytical workflows that require populating missing values under a fixed relational schema.

To overcome the limitations of existing systems, we develop \sysname, an agentic system that explores open domains to populate structured tables in Section \ref{sec:method}.
\sysname consists of two integrated core components: a web explorer and a cache manager. 
We design a novel \algname algorithm as the exploration workflow for the web explorer to conduct deep, comprehensive, and domain-specialized navigation of the open web. The cache manager stores and reuses discovered sources and navigated search paths to reduce redundant explorations and preserve coherence across related table cells.

We evaluate \sysname on \datasetname in Section \ref{sec:exp}.
Our results show that \sysname achieves 91.1\% task accuracy, outperforming the strongest baseline by approximately $2\times$ and the weakest by up to $73\times$ at the task level.
We also demonstrate \sysname’s superiority across different dimensions of \paradigmname, with particularly pronounced gains from the structured database perspective ($\mathcal{E}_2$) achieving at least a $16.6\times$ improvement.

In summary, our contributions are threefold:

\begin{enumerate}
    \item We formalize the \paradigmname task, which materializes open web data into structured databases for downstream analysis.
    
    \item We construct \datasetname, a benchmark that quantifies the challenges of \paradigmname and systematically reveals the limitations of existing AI agents.
    
    \item We develop \sysname, an agentic system that performs in-depth and comprehensive web exploration through our proposed \algname algorithm and manages navigation sources and paths in a principled manner for scalable exploration, achieving over 90\% accuracy on \datasetname.
\end{enumerate}
\section{\paradigmname}
\label{sec:usecase}
In this section, we first introduce real-world applications, i.e., the motivations for \paradigmname in Section \ref{subsec:bg}, and then formally define the task in Section \ref{subsec:problem_definition}.

\subsection{Background and Use Cases} 
\label{subsec:bg}

Researchers in scientific domains decompose their overall research objectives into analytical subproblems. At each stage, they start with precise analytical questions and an ideal database schema, but must first collect and aggregate data from a wide range of web sources into structured tables before any analysis can be conducted. 
This is particularly common in fields such as demographics \cite{Berghammer_51_34, repec:dem:demres:v:51:y:2024:i:37,Treleaven_51_32, Tsui_50_45, Moretti_50_42, Raz-Yurovich_50_34, Kan_50_21, Muniz_50_17, Vermeulen_49_20, Mogi_49_16}, economics \cite{aier_mlf_bad_behavior,aier_work_vs_welfare_tradeoff,aier_mount_laurel_synthetic_control}, sports \cite{bicudo2025comparison,cameron2025,santos2023,zhang2023, Bellinger2021QuantifyingTA, coyne2021training, costa2021}, and many other data-driven fields that use authoritative statistics published online by government agencies or international organizations. However, the required data are often distributed across deeply nested webpages, multiple reports, or separate annual releases, requiring substantial manual effort to collect and organize.

For example, Table~1 in \citet{Treleaven_51_32} presents demographic and health survey (DHS) statistics from 2000 to 2020 across multiple sub-Saharan African countries. The table uses a relational structure with \texttt{country} as the primary key, where each row corresponds to a specific country and each column represents demographic attributes, including the number of surveys conducted, survey years, the number of children under five, and the number of women of reproductive age. 
To construct this table, a key step is to populate the attributes for each country by systematically exploring the DHS website.\footnote{\url{https://www.dhsprogram.com/}} The homepage contains diverse and heterogeneous content, including announcements, reports, and multiple navigation tabs (e.g., \emph{Countries}, \emph{Data}, \emph{Publications}, \emph{Methodology}, \emph{Research}, and \emph{Topics}), many of which appear potentially relevant. Selecting the correct navigation path requires reasoning about the intended table schema: specifically, recognizing that the primary key is \emph{country} and first navigating to the \emph{Countries} section\footnote{\url{https://www.dhsprogram.com/Countries/}}, where individual country pages are listed. From there, the researcher selects the country to investigate, e.g., Benin.\footnote{\url{https://www.dhsprogram.com/countries/Country-Main.cfm?ctry_id=52}}

Similarly, Table~2 in \citet{santos2023entry} uses swim events from the 2022 FINA World Championships as the primary keys, where each row corresponds to a specific event and the columns represent swimmer statistics across different competition stages (e.g., top 16, semi-finalists, and finalists). 
The data collection procedure is structurally similar to the previous example, requiring researchers to explore the official FINA website.\footnote{\url{https://www.worldaquatics.com/}} However, in this case, different attributes for the same primary key are located on separate webpages, introducing additional complexity. 
For example, the statistics for final round\footnote{\url{https://www.worldaquatics.com/competitions/2894/16th-fina-world-swimming-championships-25m-2022/results?event=4e32ebf1-b383-4acf-87b6-8f2349da6b33&unit=final}} of the event \emph{Women 50m Freestyle} are located at a different page than the semi-finalists of the same event.\footnote{\url{https://www.worldaquatics.com/competitions/2894/16th-fina-world-swimming-championships-25m-2022/results?event=4e32ebf1-b383-4acf-87b6-8f2349da6b33&unit=semifinals-summary}} 

These examples illustrate that populating such tables requires \textbf{in-depth exploration of the open web}, involving complex and multi-step navigation.

Importantly, researchers do not restart the entire process for every primary key. In the DHS example, after collecting data for Benin and proceeding to the next country, Burundi, they do not return to the homepage. Instead, they navigate back to the \emph{Countries} section and directly select Burundi’s page\footnote{\url{https://www.dhsprogram.com/countries/Country-Main.cfm?ctry_id=51}} from there. In other words, they \textbf{use structural features of the webpages and previously discovered navigation paths to optimize their explorations}.

Finally, because the collected data are intended for downstream analysis, researchers must \textbf{ensure the table has consistent and coherent formats}. For example, time measurements extracted from the FINA website must be normalized into a unified unit, such as seconds, and displayed in a consistent \texttt{XX:XX} format.

Together, these examples reveal a recurring workflow: researchers begin with a target schema aligned with their analytical queries, but the corresponding values are scattered across multiple web sources. Apart from the analytical computation itself, a substantial portion of the effort also lies in discovering, extracting, normalizing, and materializing the required data into a structured database instance. This gap between analytical intent and fragmented, deep web data motivates the \paradigmname problem.

\subsection{Problem Definition} \label{subsec:problem_definition}
We formally define the \paradigmname problem as follows:

\textbf{Input:}
(1) an analytical query $q$ in natural language,
(2) a base domain $u$ (e.g., a website or domain scope like the DHS and FINA homepage in the examples in Section \ref{subsec:bg}), and
(3) a target relational schema $\mathcal{T}$ specifying a designated primary key and a set of attributes, together with a set of primary key values defining the rows to be populated.

\textbf{Task:}
Discover and structure relevant information from open-domain web sources within domain $u$ to populate the schema $\mathcal{T}$.

\textbf{Output:}
A materialized table instance $\mathcal{I}$ conforming to schema $\mathcal{T}$, where each cell value $\mathcal{I}[r,c]$ is instantiated with values extracted from the open web.

\section{\datasetname}
\label{sec:data}

\begin{table*}[t]
\centering
\small
\caption{Statistics of \datasetname.}
\label{tab:dataset_stats}
\resizebox{0.8\textwidth}{!}{
\begin{tabular}{lccccccc}
\toprule
\textbf{Domain} & \# Surveys & \# Tables & \# Base Domains & \# Primary Keys & \# Cols & \# Cells & Avg. \# Search Depth \\
\midrule
Demographics & 8 & 8 & 7 & 29 & 49 & 188 & 3.75 \\
Sports & 7 & 30 & 9 & 162 & 184 & 885 & 4.35 \\
Finance & 11 & 22 & 10 & 99 & 85 & 379 & 3.33 \\
Economics & 5 & 12 & 6 & 46 & 41 & 154 & 3.39 \\
Food & 9 & 19 & 9 & 75 & 90 & 354 & 3.12 \\
Climate & 8 & 14 & 8 & 47 & 57 & 189 & 3.08 \\
\midrule
\textbf{Total} & \textbf{48} & \textbf{105} & \textbf{49} & \textbf{458} & \textbf{506} & \textbf{2149} & \textbf{3.81} \\
\bottomrule
\end{tabular}
}
\end{table*}

Existing benchmarks do not capture the requirements of \paradigmname tasks. 
Traditional table-filling benchmarks \cite{lakefill}, information retrieval benchmarks \cite{yang2024crag, thakur2021beirheterogenousbenchmarkzeroshot}, data science benchmarks \cite{lai2025kramabenchbenchmarkaisystems, Hu_2024}, and NL2SQL benchmarks \cite{lei2024spider, li2024can} assume that relevant documents or databases are locally available and well-structured. 
Existing open-domain search benchmarks either focus on retrieving simple and independent textual content \cite{he2024webvoyager} or assume that structured databases have already been instantiated online \cite{Hu_2025}. 

To systematically evaluate the \paradigmname problem in real-world settings, we construct \datasetname, a benchmark derived from published academic papers across diverse scientific domains.

To ensure that \datasetname reflects realistic and non-trivial data collection tasks, we apply the following selection criteria:

\begin{enumerate}
    \item The study requires data that can be structured into a relational table with unambiguous primary keys and attributes.
    \item The required data for each study must span multiple webpages (i.e., cannot be retrieved from a single page).
    \item All data must be directly extractable from webpages without complex offline processing or heavy data transformation.
    \item All relevant webpages must be publicly accessible and must not require authentication or restricted access.
\end{enumerate}

\begin{figure}[t]
    \graphicspath{{figures/}}
    \centering
    \includegraphics[width=\columnwidth]{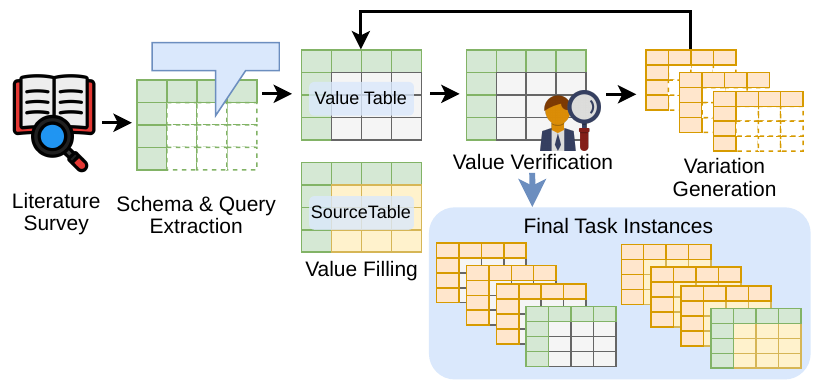}
    \caption{Data collection workflow of \datasetname.}
    \label{fig:datacollection}%
\end{figure}

As we document in Figure~\ref{fig:datacollection}, our data collection process goes through a systematic workflow composed of the following stages.

\minihead{Literature Survey}
To ensure \emph{(i)} ecological validity, we derive tasks from published research papers across 6 common domains; \emph{(ii)} recency and timeliness, we collect papers up to a unified cutoff date of December 2025; \emph{(iii)} sufficient task diversity and balanced coverage across fields, we determine domain-specific starting years based on two considerations:
(1) the publication frequency and overall volume of papers in the venue, and
(2) the proportion of recent papers that satisfy our task selection criteria.
Specifically, we examine the most recent issues of each venue to estimate the percentage of papers, denoted by $\alpha$, whose underlying data meet our eligibility requirements. Because this proportion varies substantially across domains (e.g., many sports studies rely on interview-based or restricted data, whereas economics studies mostly use publicly accessible datasets), earlier starting years are required in domains with lower $\alpha$ to obtain a sufficient number of eligible tasks. We select starting years such that approximately $10/\alpha$ papers are surveyed per domain, yielding roughly 10 qualifying tasks in expectation:

\begin{itemize}
    \item \textbf{Demographics:} Papers published from September 2023 to December 2025 in \emph{Demographic Research}.\footnote{\url{https://www.demographic-research.org/}}
    \item \textbf{Sports:} Papers published from 2021 to 2025 in \emph{Journal of Sports Science \& Medicine}.\footnote{\url{https://www.jssm.org/}}
    \item \textbf{Finance:} Papers published in 2025 in \emph{Journal of Finance}.\footnote{\url{https://onlinelibrary.wiley.com/journal/15406261}}
    \item \textbf{Economics:} Papers published in 2025 by the \emph{American Institute for Economic Research}.\footnote{\url{https://aier.org/features/paper/}}
    \item \textbf{Food:} Papers published in 2025 in \emph{Nature Food}.\footnote{\url{https://www.nature.com/natfood/}}
    \item \textbf{Climate:} Papers published from August 2025 to December 2025 in \emph{Nature Climate Change}.\footnote{\url{https://www.nature.com/nclimate/}}
\end{itemize}
Overall, we reviewed 1,004 papers and identified 48 that satisfy our task construction criteria, with the domain-wise distribution documented in Table \ref{tab:dataset_stats}.

\minihead{Schema and Query Generation}
For each reviewed paper, we construct table schema directly from tables or figures that summarize collected data (e.g., metadata tables, cross-country comparisons, or survey statistics) when such structured representations are available. 
In cases where no explicit table or figure is provided, we derive the schema by summarizing the data collection process described in the text, ensuring that the resulting relational structure satisfies our selection criteria while faithfully reflecting the underlying data analytical workflow.
 Each resulting task instance consists of:
\begin{itemize}
    \item a natural language analytical query $q$ describing the research objective of the paper,
    \item a base domain $u$ specifying the authoritative data source (i.e., the official homepage of the relevant organization or data portal);
    \item a target relational schema $\mathcal{T}$ specifying a designated primary key and a set of attributes, together with a set of primary key values defining the rows to be instantiated.
\end{itemize}

The objective is to construct a materialized table instance over $\mathcal{T}$ by systematically retrieving and structuring values from diverse open-domain web sources.

\minihead{Value Collection}
A team of two annotators is responsible for instantiating the table values using only the query, schema, and base domain (i.e., the same inputs provided to evaluated systems). Annotators independently retrieve and fill in all cell values from public web sources, recording both the materialized table and the source URLs.

\minihead{Value Verification}
To ensure correctness and reliability, a team of four annotators verifies the collected values. Verification involves comparing reconstructed tables against the original paper tables whenever possible. In cases where discrepancies arise (e.g., evolving public datasets), annotators resolve conflicts until full mutual agreement based on documented web evidence.

\minihead{Variation Generation}
To increase robustness and evaluation coverage, we generate additional variations within the same base domain and difficulty level. For example, if an original study analyzes results from a \emph{Female} Super Sevens Tournament~\cite{bicudo2025comparison}, we construct a parallel task for the corresponding \emph{male} tournament. All variations then undergo the same verification procedure, as previously described.

Following this workflow, we construct 105 tasks from 48 published studies across 6 scientific domains. We provide detailed statistics of \datasetname in Table~\ref{tab:dataset_stats}.
In total, \datasetname contains 458 primary key values and 506 attributes, resulting in 2149 cell values that must be instantiated from distributed web sources (i.e., 2149 retrieval operations are required). 
% On average, each task contains 5.4 primary key instances and 5.9 attributes, yielding roughly 32 cell values per task. 
% This structure indicates that even individual tasks require coordinated aggregation across multiple entities and attributes rather than isolated value retrieval. Systems must systematically navigate heterogeneous webpages, collect interdependent values across rows and columns, and maintain schema-level consistency throughout the construction process. 
The scale, diversity, and structural complexity of \datasetname reflect realistic database materialization workloads observed in domain research, establishing it as a rigorous and representative benchmark for evaluating the \paradigmname problem.

\section{\sysname}
\label{sec:method}

\begin{figure}[t]
    \graphicspath{{figures/}}
    \centering
    \includegraphics[width=\columnwidth]{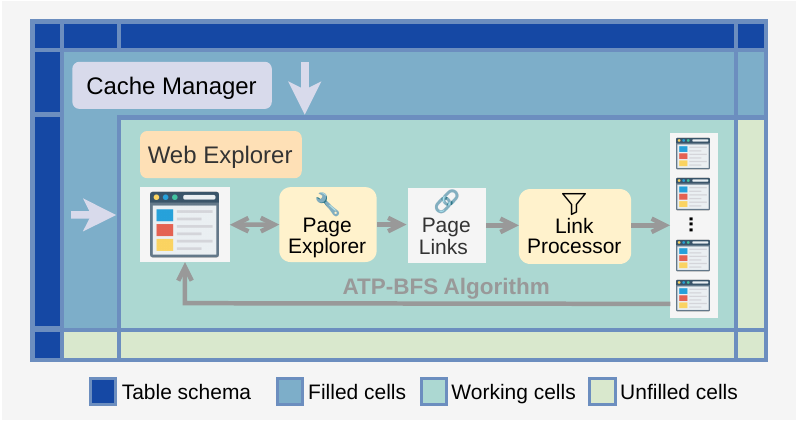}
    \caption{Overview of \sysname.}
    \label{fig:mainmethod}%
\end{figure}

In this section, we introduce \sysname, an agentic system that aggregates and structures web information. Given a query, a base domain, and an output table schema defined by its primary keys (rows) and attributes (columns), \sysname fills in each cell in sequence. As we illustrate in Figure~\ref{fig:mainmethod}, \sysname consists of two core components: (1) a web explorer (Section~\ref{subsec:web_explorer}), which performs in-depth exploration to extract cell-specific information from web pages and equips the agent with mechanisms to construct and use its own MCP tools, and (2) a cache manager (Section~\ref{subsec:domain_cache}), which caches and reuses previously discovered sources and navigation paths to make use of the structural dependencies among cells. To populate a target cell, the cache manager first queries neighboring page caches. If they do not contain the required information, it retrieves and generates a ranked list of candidate webpages from the path cache. The web explorer then executes the \algname traversal algorithm from these selected initial webpages in order.

Conceptually, \sysname mirrors a processor-memory co-design: the web explorer functions as the core processing unit, executing the active exploration logic for each cell, while the cache manager serves as the storage unit, persistently maintaining previously discovered sources and structural paths to reduce redundant computation and improve coherence across cells. The synergistic interaction between these two components enables comprehensive and efficient collection and aggregation of data from the open web.

\subsection{Web Explorer and the \algname Algorithm} 
\label{subsec:web_explorer}
We introduce the \textit{augment-then-prune} breadth-first web exploration algorithm (\algname), an agentic web exploration algorithm (Algorithm~\ref{alg:atp_bfs}) that serves as the backbone of the web explorer by conducting deep and comprehensive exploration on all visited webpages.
Similar to classic BFS, \algname starts from a root webpage and iteratively expands a frontier queue in a layer-by-layer manner. For each visited page, outgoing links to unvisited webpages are added to the frontier until the system identifies and returns the target cell value.

As we illustrate in Figure~\ref{fig:mainmethod}, the web explorer executes \algname traversal with agentic interactions that interpret, interact with, and extract information from webpages during navigation.

We focus on two key innovations in the web explorer: (1) how it interacts with webpages to extract links and obtain answers (page explorer), and (2) how it determines the next layer of candidate links to explore through the \textit{augment then prune} process (link processor).

\minihead{Page Explorer}
For each visited webpage $u$, the web explorer performs page-wise exploration and applies a decision function $\Phi(u)$ through the page explorer as a subagent:
\[
\Phi(u) \in \{\texttt{Answer}(v),\ \texttt{Proceed}(L),\ \texttt{Stop}\},
\]
where:
\begin{itemize}
    \item $\textsc{Answer}(v)$ returns a value $v$ for the target cell if the required information is present on $u$;
    \item $\textsc{Proceed}(L)$ returns a set of outgoing links $L$ discovered on $u$ for further exploration;
    \item $\textsc{Stop}$ indicates that neither a valid answer nor useful links are found.
\end{itemize}

In practice, webpages differ substantially in how content is presented and can be categorized into three common classes: \textit{static pages}, where all relevant content is fully revealed in the HTML source, \textit{dynamically loaded pages}, where critical information may be hidden behind client-side rendering, JavaScript-driven UI components, or interactive elements such as tabs, accordions, pagination controls, and/or filters, and \textit{online documents} (e.g., PDFs or images such as PNG/JPG/WebP). 

When the web explorer encounters an online document URL, it invokes a file inspector that downloads the file for visual inspection. For image files, the inspector directly analyzes the image. For PDFs, it renders pages to images and examines them sequentially. After inspecting each image, the inspector decides whether the target cell value is already visible; if not, whether the answer is likely to appear on another page of the same file (and thus exploration should continue within the file), or whether further file exploration is unlikely to succeed, enabling early termination.

For webpages, the web explorer next classifies whether the page can be reliably treated as a static webpage or requires dynamic interactions. To do so, it compare two textual views extracted from the same webpage: a higher-fidelity observation obtained after rendering (which includes dynamically loaded or UI-revealed contents), and a lower-fidelity extract derived from static HTML. The page is treated as dynamic if the static extract is considered to be incomplete for the target cell. Specifically, a page is classified as dynamic if any of the following conditions hold: (i) the higher-fidelity observation contains concrete, query-relevant facts such as numbers, dates, names, or table rows that are missing from the static extract; (ii) the higher-fidelity observation reveals structural cues such as section headers, tabs, pagination indicators, or navigation items whose corresponding content is absent from the static extract; (iii) the static extract appears truncated, boilerplate, or substantially less informative than the rendered observation; or (iv) the static extract itself exposes UI controls (e.g., tabs, dropdowns, filters, or ``load more'' elements) without displaying their associated content. 
% Only when none of these conditions are met do we treat the page as static.

If the page is classified as static, the page explorer directly extracts its visible text and outgoing links from the static HTML component, and uses these signals together with a rendered screenshot to decide the return value among the three outcomes as we previously defined. If the page is classified as dynamic, the page explorer switches to an interactive exploration mode in which it incrementally reveals hidden contents. 

Specifically, for dynamic webpages, the page explorer operates in a multi-turn manner. In each turn, it is provided with the webpage’s JavaScript accessibility structure and textual observations. It begins with an initial accessibility snapshot that includes common interactive components such as buttons and tabs.
At each turn, the explorer performs one of the following 5 page-local actions:

\begin{enumerate}
    \item \texttt{update\_accessibility}: this operation allows the web explorer to dynamically extend its interaction interface by identifying specialized UI elements (e.g., custom-styled tabs or buttons) and registering them as clickable components. In effect, it enables the agent to design its own MCP tools by constructing page-specific interaction primitives.
    \item \texttt{click}: this operation enables the agent to interact with one or more UI elements (e.g., expand/collapse, ``show more'', or tab switches) to reveal additional on-page content.
    \item \texttt{answer}: this operation terminates page-wise exploration when the required cell value becomes visible on the current page. It corresponds to the \texttt{Answer} outcome.
    \item \texttt{extract\_links}: this operation surfaces all visible hyperlinks for continued exploration when the answer is not found on the current page but the page may lead to relevant sources. It corresponds to the \texttt{Proceed} outcome.
    \item \texttt{stop}: this operation terminates exploration when neither the answer nor useful outgoing links are present on the current page.
    It corresponds to the \texttt{Stop} outcome.
\end{enumerate}

The web explorer is guided to fully apply (1) and (2) to reveal all available on-page contents before executing any of (3), (4), or (5).

During page exploration, the web explorer uses existing information as contextual signals to support its decision-making. Conversely, when it encounters clear and unambiguous evidence that contradicts earlier entries, it proactively revises the affected cell values to maintain consistency. 
This self-correction mechanism improves the robustness of \sysname.
% By exploiting these structural dependencies across table cells, the explorer improves robustness and achieves higher overall accuracy. This correction mechanism further facilitates the organization of collected information into consistent formats suitable for downstream computation.

\minihead{Link processor}
If we were to follow the traditional BFS algorithm, the web explorer would (1) enqueue all outgoing links on each visited page, (2) explore only the links generated from the previous layers, and (3) traverse them strictly in the order they are added to the queue. Such a strategy is inefficient for web exploration for several reasons. First, a single webpage can contain a large number of outgoing links, causing the frontier size to grow rapidly and exponentially with depth. Second, the information required to answer a query, especially dated or archival content, may reside deep in the website hierarchy rather than within a few navigation steps from the root. Third, outgoing links on a webpage are not ordered by relevance to the query, which further delays reaching the target page and exacerbates the inefficiency caused by the large branching factor.

To address these challenges, we develop a link processor that operationalizes the \textit{augment-then-prune} strategy in \algname. 
Given the extracted outgoing links together with their annotation texts, the link processor produces a compact, relevance-oriented frontier for the next BFS layer. Specifically, it performs three stages: \textit{Augment}, \textit{Select}, and \textit{Rank}.

\begin{enumerate}
\item \textit{Augment (pattern-based URL proposal).}
The outgoing links on a webpage are often incomplete for the current query target (e.g., a page contains links for the year 2024, but the query requires year 2025). In the augment stage, we propose additional candidate URLs by inferring local URL patterns from the observed links and anchor texts. Specifically, the link processor analyzes the set of links listed on the webpage for recurring templates such as path segments (e.g., \texttt{/results/2024/}), slugs, identifiers, pagination markers, or query-string conventions. It then generates new URLs via minimal edits to existing on-page links, which preserves the site’s structure and avoids unrealistic hallucinations. A minimal edit changes at most one path segment (or applies a uniform multi-segment substitution such as \texttt{2024}$\rightarrow$\texttt{2025} across all occurrences), and does not introduce new query parameters or additional path depth. In rare cases where the observed links imply multiple closely related domains (e.g., sibling subdomains \texttt{\url{liheap-fy24-data-dashboard-hhs-acf.hub.arcgis.com}} and \texttt{\url{liheap-fy25-data-dashboard-hhs-acf.hub.arcgis.com}}), the link processor performs a domain edit while keeping the full path unchanged.

\item \textit{Select (joint pruning under a budget).}
After augmentation, we obtain a joint pool consisting of (i) the original outgoing links extracted from the current page and (ii) the newly proposed links from the augment stage. Exploring all of these links is impractical due to the large branching factor, so the select stage prunes this pool to a budget of at most $K$ links (user-configurable). Selection is based on the textual descriptions available for each URL, including anchor text, surrounding snippet text, and URL tokens (e.g., year segments, team names, keywords). 
% Importantly, selection is joint across original and augmented links, so that augmentation does not dominate the frontier and the explorer still retains coverage of organically provided navigation paths. 
The link processor is guided to reserve a reasonable portion of the $K$ budget for original links and the remainder for augmented links, preventing the frontier from collapsing to only augmented URLs or only on-page links.

\item \textit{Rank (query and structure aware ordering).}
Finally, the selected links are ranked to determine the exploration order for the next BFS layer. Ranking is conditioned on (i) the user query, (ii) the target cell specification (the column name and row identifier), and (iii) any already collected evidence. The link processor prioritizes URLs whose textual descriptions suggest they directly contain the target value, as well as URLs that likely serve as intermediate navigation hubs toward relevant content.

\end{enumerate}

The final output of the link processor after the three stages is a ranked list of up to $K$ URLs that becomes the next-layer frontier for \algname.

\begin{algorithm}[t]
\caption{\algname Agentic Web Exploration}
\label{alg:atp_bfs}
\small
\KwIn{Query $q$, target cell $(r,c)$, root URL $u_0$, explore width $K$}
\KwOut{Cell value or $\emptyset$}

Initialize frontier queue $Q \leftarrow [u_0]$\;
Initialize visited set $V \leftarrow \emptyset$\;

\While{$Q \neq \emptyset$}{
    $u \leftarrow$ Dequeue$(Q)$\;
    \If{$u \in V$}{continue}
    $V \leftarrow V \cup \{u\}$\;

    $page \leftarrow$ Fetch$(u)$\;

    \tcp{Page-wise exploration}
    $result \leftarrow$ PageExplorer$(page, q, (r,c))$\;

    \If{$result = \texttt{Answer}(v)$}{
        \Return{$v$}
    }
    \If{$result = \texttt{Proceed}(L)$}{
        \tcp{Link augment-then-prune}
        $L_{next} \leftarrow$ LinkProcessor$(L, q, (r,c), K)$\;

        \ForEach{$u' \in L_{next}$}{
            \If{$u' \notin V$}{
                Enqueue$(Q, u')$\;
            }
        }
    }
    
    \tcp{else $result = \texttt{Stop}$}
}
\Return{$\emptyset$}
\end{algorithm}

\subsection{Cache Manager}
\label{subsec:domain_cache}

While the web explorer focuses on breadth-first exploration to ensure coverage, the cache manager complements it by exploiting depth-wise regularities in website structures, resembling a constrained depth-first search over previously successful navigation paths. The key insight is that neighboring cells within the same table, as well as neighboring webpages within the same domain, exhibit strong structural dependencies: their values are often located on the same pages or along highly similar navigation paths.

For each successfully filled cell, \sysname caches both the final source page and the full exploration path from the initial root to that page. Caching the entire path is crucial because intermediate pages often encode structural cues (e.g., year indices, entity lists, or category hubs) that generalize across neighboring cells. 
% In addition, we record automatic URL redirections encountered during exploration, ensuring that future lookups operate on canonical URLs rather than obsolete or transient ones.

\sysname's cache is organized into two levels, both indexed by primary key and column. The \textit{page-level cache} records mappings from the source webpages to the extracted values, enabling immediate reuse when multiple cell values are populated on the same page. The \textit{path-level cache} stores full exploration paths, allowing the system to reuse deeper navigation patterns when structurally similar cells are encountered.

Before invoking the web explorer for a target cell, \sysname first consults the cache manager. It checks whether the source page used by the upper or left neighboring cell already contains the required information. If not, \sysname retrieves the cached exploration paths from these neighboring cells and performs a path search to identify promising next pages to visit. Specifically, given the two cached paths, \sysname ranks a set of candidate URLs of size at most $K$ by prioritizing URLs whose patterns suggest relevance to the target primary key and column, and those located near likely divergence points where navigation paths for different cells begin to differ. When appropriate, new candidate URLs are proposed through minimal edits to existing path URLs, such as substituting year or identifier segments, while preserving the observed site structure, following the same principle as the link processor.

The resulting ranked candidate URLs (or an empty set for the first cell with no cache), together with the base domain, are then used as root nodes (i.e., the $u_0$ in Algorithm \ref{alg:atp_bfs}) for the web explorer. This cache-guided initialization not only reduces redundant exploration but also extends \sysname's search depth.

\vspace{.5em}
Overall, the architecture of \sysname enables more comprehensive and in-depth web exploration than existing web search systems by systematically diving into deep and archival pages when necessary, while simultaneously reducing exploration cost through effective reuse of cached pages and navigation paths.

\section{Experiments}
\label{sec:exp}

We first describe the experimental setup in Section~\ref{subsec:exp_setup}. 
We then present the system-level performance of all agents on \datasetname in Section~\ref{subsec:overall_res}, including a detailed analysis across the three dimensions ($\mathcal{C}_0$, $\mathcal{E}_1$, and $\mathcal{E}_2$) for solving \paradigmname.
To better understand the mechanisms underlying \sysname's performance, we provide component analyses of the web explorer in Section~\ref{subsec:exp_web_explorer} and the cache manager in Section~\ref{subsec:exp_cache}. 
Finally, in Section~\ref{subsec:exp_ablation}, we conduct ablation studies to evaluate the standalone performance of the web explorer and the incremental benefit of the cache manager.
% demonstrate the critical role of $\mathcal{E}1$ and $\mathcal{E}2$.

\subsection{Experiment Setup}
\label{subsec:exp_setup}
We evaluate the performance of \sysname and 6 baseline agents on \datasetname.
\subsubsection{Baselines}
We select the following 6 state-of-the-art agents with advanced web search capabilities as baselines.

\minihead{AG2 \cite{ag2_docs}}
AG2 is a modular agent framework that coordinates multiple specialized agents (for example, a planner, researcher, and executor) to solve tasks via iterative reasoning and tool use, including web search and information gathering. It emphasizes flexible agent composition and structured collaboration to improve reliability on multi-step problems.

\minihead{AutoGPT \cite{autogpt_repo}}
AutoGPT is an autonomous agent that decomposes a high-level goal into subgoals, repeatedly plans, searches the web, and executes actions to make progress with minimal human intervention. It maintains short-term and long-term context (for example, via memory or logs) to support iterative refinement over long horizons.

\minihead{AutoGen \cite{microsoft_autogen}}
AutoGen is a multi-agent conversation framework where agents communicate through structured messages to jointly plan, retrieve information, and complete tasks. It supports tool-using agents (such as web browsing, code execution, and function calls) and is designed to make complex workflows easier to orchestrate and evaluate.

\minihead{OpenAI ResearchBot \cite{openai2024researchbot}}
The OpenAI ResearchBot is a research-focused agent designed to perform multi-step online investigation, including querying the web, reading sources, synthesizing evidence, and producing grounded summaries.

\minihead{Open Deep Research \cite{opendeepresearch}}
Open Deep Research is an open-source reproduction of a ``deep research'' \cite{google_deep_research_2024} style agent that conducts iterative web exploration, note-taking, and synthesis across a wide range of sources to answer complex queries. It follows a structured research workflow that alternates between planning, searching, reading, and summarizing, aiming for comprehensive coverage and traceable reasoning.

\minihead{WebVoyager \cite{he2024webvoyager}}
WebVoyager is an end-to-end web navigation agent that interacts with real websites by following links, clicking UI elements, and extracting information from pages to complete tasks. It couples high-level planning with low-level browser actions, enabling it to handle workflows that require multi-page navigation and on-page interaction.

\vspace{.5em}
All agents operate under a cell-wise extraction setting, where each task run corresponds to filling a single table cell. For each run, agents are provided with already filled values as contextual information. All agents are prompted using \texttt{gpt-5-2025-08-07}~\cite{openai_gpt5_api}. We set the exploration width $K=10$ for \sysname.

\subsubsection{Evaluation Metrics}

% We begin evaluation at the \textbf{cell level}, since each \paradigmname task requires correctly retrieving and organizing multiple structured values.

We first define the cell matching criterion that underlies all value-based comparisons. We then introduce our primary metric, followed by metrics tailored to different dimensions. Here, $R$ and $C$ denote the number of rows and columns in each table, and $i$ and $j$ represent row and column indexes.

\minihead{Cell Matching Criteria}
For each predicted cell $\hat{y}_{i,j}$ and ground-truth cell $y_{i,j}$, we adopt two complementary matching strategies:

\begin{enumerate}
    \item \textbf{Exact Match.}
    \[
    \mathbb{I}_{\text{match}}(i,j) =
    \begin{cases}
    1 & \text{if } \hat{y}_{i,j} = y_{i,j}, \\
    0 & \text{otherwise}.
    \end{cases}
    \]
    This metric is strict and fully reproducible.

    \item \textbf{LLM-as-a-Judge.}
    Exact match can be overly strict when formatting differs, but semantics are equivalent (e.g., ``1\%'' vs.\ ``0.01'', rounding differences, or alternative naming conventions). 
    To account for such cases, we use GPT-4o \cite{openai_gpt4o_api} as a semantic judge that outputs a binary decision following data agent evaluation traditions \cite{xu2026inside}:
    \[
    \mathbb{I}_{\text{llm}}(i,j) \in \{0,1\}.
    \]
\end{enumerate}

\minihead{Primary Metric: Task-Level Accuracy}
Our primary metric evaluates the success rate of structured data construction at the \emph{task level}.
We compute the task accuracy as the percentage of correctly filled cells in the materialized table:

\[
\text{TaskAcc} =
\frac{
\sum_{i=1}^{R} \sum_{j=1}^{C}
\mathbb{I}(i,j)
}{
R C
}.
\]

We then average across \emph{all tasks in \datasetname}.
This metric directly reflects the overall success rate of structured data construction, capturing the integrated effectiveness of $\mathcal{C}0$, $\mathcal{E}1$, and $\mathcal{E}2$.

\minihead{Dimension-Level Metrics} Beyond the primary task-level metric, we introduce dimension-level metrics to disentangle performance across $\mathcal{C}0$, $\mathcal{E}1$, and $\mathcal{E}2$.

\textbf{($\mathcal{C}_0$) In-Depth Information Discovery.}
To assess fine-grained retrieval performance, we compute CellAcc as the average accuracy across \emph{all cells in \datasetname}. This metric reflects the overall proportion of correctly retrieved values and directly captures the system’s ability to perform in-depth information discovery ($\mathcal{C}_0$).

\textbf{($\mathcal{E}1$) Structural Correlation Identification.}
To evaluate whether systems correctly identify and exploit structural dependencies across table cells, we measure full row and full column accuracy.

\emph{Full Row Accuracy:}
\[
\text{RowAcc}_i =
\mathbb{I}
\left(
\sum_{j=1}^{C} \mathbb{I}(i,j) = C
\right).
\]

\emph{Full Column Accuracy:}
\[
\text{RowAcc}_j =
\mathbb{I}
\left(
\sum_{i=1}^{R} \mathbb{I}(i,j) = R
\right).
\]

The row accuracy is averaged across \emph{all rows in \datasetname}, and the column accuracy is averaged across \emph{all columns in \datasetname}.
These metrics reflect whether the system can consistently apply structural patterns and reuse navigation strategies across related entries, directly measuring $\mathcal{E}1$.

\textbf{($\mathcal{E}2$) Structured Database Construction.}
We define table-level accuracy as the percentage of fully correct tables:
\[
\text{TableAcc} =
\mathbb{I}
\left(
\sum_{i=1}^{R} \sum_{j=1}^{C}
\mathbb{I}(i,j) = R C
\right).
\]

The table accuracy is averaged across \emph{all tasks in \datasetname}. This metric evaluates whether the system successfully constructs a complete, coherent, and queryable database instance, directly reflecting $\mathcal{E}2$.

\paragraph{Metric Hierarchy.}
These metrics form a natural hierarchy:
\[
\text{CellAcc} \;\ge\; \text{RowAcc}, \text{ColAcc} \;\ge\; \text{TableAcc}.
\]
Conceptually, the hierarchy mirrors the progression from local value retrieval to global relational integrity: Cell accuracy measures retrieval correctness ($\mathcal{C}0$), structural accuracy measures consistency across related entries ($\mathcal{E}1$), and table accuracy measures complete and coherent database construction ($\mathcal{E}2$).

\subsection{\sysname achieves promising performance on \paradigmname tasks}
\label{subsec:overall_res}
\begin{table*}[t]
\centering
\small

\caption{Accuracy ($\uparrow$) of different agents on \datasetname in percentages (\%). GPT-4o columns refer to LLM-as-a-judge evaluations, and Match columns refer to exact match evaluations.
The optimal values for each metric are highlighted with both bold and underline formatting. We report the rank of each agent under each metric in italicized parentheses.}
\label{tab:overall_res}
\renewcommand{\arraystretch}{1.4}
\setlength{\dashlinegap}{0pt}
\setlength{\belowrulesep}{0pt}
\setlength{\aboverulesep}{0pt}
\setlength{\arrayrulewidth}{0.6pt}
\resizebox{\textwidth}{!}{
\begin{tabular}{l|cc|cc|cc|cc|cc}
\firsthline
\multicolumn{1}{c|}{\multirow{2}{*}{\textbf{System}}}
& \multicolumn{2}{c|}{Cell Acc. ($\mathcal{C}0$)}
& \multicolumn{2}{c|}{Row Acc. ($\mathcal{E}1$)}
& \multicolumn{2}{c|}{Column Acc. ($\mathcal{E}1$)}
& \multicolumn{2}{c|}{Table Acc. ($\mathcal{E}2$)}
& \multicolumn{2}{c}{\textbf{Task Acc. ($\mathcal{C}0$, $\mathcal{E}1$, $\mathcal{E}2$)}}
\\
\cmidrule(lr){2-3}\cmidrule(lr){4-5}\cmidrule(lr){6-7}\cmidrule(lr){8-9}\cmidrule(lr){10-11}
& GPT-4o & Match
& GPT-4o & Match
& GPT-4o & Match
& GPT-4o & Match
& \textbf{GPT-4o} & \textbf{Match} \\
\specialrule{0.05em}{0pt}{0pt}
AG2  & 45.51 \emph{(2)} & 36.39 \emph{(2)} & 15.10 \emph{(2)} & 8.17 \emph{(2)} & 18.69 \emph{(2)} & 15.61 \emph{(2)} & 1.90 \emph{(3)} & 0.95 \emph{(2)} & 46.48 \emph{(2)} & 35.90 \emph{(2)} \\
AutoGPT & 35.74 \emph{(3)} & 29.32 \emph{(3)} & 10.89 \emph{(3)} & 5.22 \emph{(3)} & 10.26 \emph{(4)} & 7.44 \emph{(4)} & 1.90 \emph{(3)} & 0.95 \emph{(2)} & 34.73 \emph{(3)} & 28.42 \emph{(3)} \\
AutoGen & 31.32 \emph{(4)} & 26.57 \emph{(4)} & 7.84 \emph{(4)} & 3.33 \emph{(4)} & 13.71 \emph{(3)} & 11.21 \emph{(3)} & 2.86 \emph{(2)} & 0.95 \emph{(2)} & 29.54 \emph{(4)} & 24.70 \emph{(4)} \\
OpenAI ResearchBot & 16.71 \emph{(5)} & 5.21 \emph{(5)} & 1.56 \emph{(5)} & 0.43 \emph{(5)} & 3.84 \emph{(5)} & 0.61 \emph{(5)} & 0.00 \emph{(5)} & 0.00 \emph{(5)} & 16.61 \emph{(5)} & 4.85 \emph{(5)} \\
WebVoyager & 2.05 \emph{(6)} & 1.12 \emph{(6)} & 0.24 \emph{(6)} & 0.00 \emph{(6)} & 0.61 \emph{(6)} & 0.14 \emph{(6)} & 0.00 \emph{(5)} & 0.00 \emph{(5)} & 2.93 \emph{(6)} & 1.54 \emph{(6)} \\
Open Deep Research & 1.35 \emph{(7)} & 0.51 \emph{(7)} & 0.00 \emph{(7)} & 0.00 \emph{(6)} & 0.00 \emph{(7)} & 0.00 \emph{(7)} & 0.00 \emph{(5)} & 0.00 \emph{(5)} & 1.24 \emph{(7)} & 0.35 \emph{(7)} \\
\specialrule{0.05em}{0pt}{0pt}
\textbf{\sysname} & \textbf{\underline{91.76}} \emph{(1)} & \textbf{\underline{67.24}} \emph{(1)} & \textbf{\underline{75.75}} \emph{(1)} & \textbf{\underline{39.68}} \emph{(1)} & \textbf{\underline{78.30}} \emph{(1)} & \textbf{\underline{52.30}} \emph{(1)} & \textbf{\underline{52.38}} \emph{(1)} & \textbf{\underline{23.81}} \emph{(1)} & \textbf{\underline{91.07}} \emph{(1)} & \textbf{\underline{69.48}} \emph{(1)} \\
\addlinespace[0.5pt]
\hdashline[3pt/3pt]
\addlinespace[0pt]
\sysname's Web Explorer & 86.64 & 60.91 & 61.07 & 27.65 & 64.25 & 38.73 & 31.43 & 11.43 & 84.37 & 60.04 \\
\lasthline
\end{tabular}
}
\end{table*}

We report the overall performance of \sysname in Table \ref{tab:overall_res}. We have the following analysis and conclusions:

\minihead{\sysname outperforms SOTA agents}
\sysname significantly outperforms all baselines across every \paradigmname dimension and evaluation setting. Under LLM-as-a-judge evaluation, \sysname achieves 91.1\% task accuracy, nearly doubling AG2 and exceeding AutoGPT and AutoGen by more than $2.6\times$ and $3.1\times$, respectively. Compared to OpenAI ResearchBot, WebVoyager, and Open Deep Research, the margin becomes even more substantial, reaching up to $73\times$ over Open Deep Research.
This dominance is consistent under exact match evaluation, where \sysname achieves 69.5\% task accuracy: under the stricter exact-match metric, \sysname consistently achieves approximately $1.9\times$ the performance of AG2 and over $2.4\times$ that of AutoGen.

For cell accuracy ($\mathcal{C}0$), \sysname reaches 91.8\% under the LLM-as-a-judge metric, which is more than $2\times$ AG2 and nearly $3\times$ AutoGen. It further substantially surpasses AutoGPT and OpenAI ResearchBot, achieving over a $5\times$ improvement compared to the latter. This demonstrates that \sysname is highly reliable at fine-grained information discovery.

For row and column accuracy ($\mathcal{E}1$), the gap widens further. On row accuracy under the LLM-as-a-judge metric, \sysname achieves 75.8\%, which is $5.0\times$ AG2, $6.9\times$ AutoGPT, and nearly $10\times$ AutoGen. On column accuracy under the LLM-as-a-judge metric, \sysname reaches 78.30\%, over $4.2\times$ AG2 and more than $7.6\times$ AutoGPT. These results indicate that \sysname effectively identifies and leverages structural correlations across cells.

For table accuracy ($\mathcal{E}2$), which requires globally consistent reasoning across the entire table, \sysname achieves 52.4\% using the LLM-as-a-judge metric. This corresponds to a $27\times$ improvement over AG2 and an $18\times$ improvement over AutoGen. Under exact match, \sysname achieves 23.8\%, whereas all major baselines remain at or below 0.95\%. This large gap demonstrates that most existing agents fail to maintain global consistency across rows and columns, while \sysname successfully integrates $\mathcal{C}0$ and $\mathcal{E}1$ with coherent table-level reasoning ($\mathcal{E}2$).

The large performance gaps at $\mathcal{E}1$ and $\mathcal{E}2$ dimensions underscore that structured multi-level reasoning, rather than isolated retrieval, is the key bottleneck in \paradigmname tasks. Overall, these results demonstrate that existing web agents struggle with coordinated structural reasoning, whereas \sysname successfully integrates in-depth information discovery ($\mathcal{C}0$), structural dependency modeling ($\mathcal{E}1$), and global table consistency ($\mathcal{E}2$) into a unified and scalable system.

\minihead{Consistent, robust, and complementary evaluation metrics enable reliable and discriminative comparisons on \paradigmname tasks}
Within the same accuracy level, the trends between exact match and LLM-as-a-judge are highly consistent in ranking, confirming the robustness of our conclusions. The LLM-based evaluation typically yields slightly higher scores because it tolerates semantically equivalent answers that differ in surface form.

More importantly, LLM-as-a-judge provides finer granularity when accuracy is low, enlarging performance gaps that exact match cannot distinguish. For example, both WebVoyager and Open Deep Research have 0\% row accuracy under exact match. However, under LLM evaluation, WebVoyager achieves 0.24\% while Open Deep Research remains at 0\%, indicating that WebVoyager exhibits stronger $\mathcal{E}1$ than Open Deep Research. Similarly, for table accuracy, AutoGen achieves 2.86\% under LLM evaluation, outperforming AG2 and AutoGPT (both 1.90\%), while exact match reports identical scores (0.95\%), masking this difference.

From the value perspective, across evaluation levels, we observe strong correlations between the three dimensions and overall task accuracy. The closeness between task and cell accuracy indicates that strong fine-grained retrieval ($\mathcal{C}_0$) forms a necessary foundation for end-to-end success.
The importance of structural extensions ($\mathcal{E}_1$ and $\mathcal{E}_2$) becomes evident when moving from cell-level to row-, column-, and table-level metrics. Systems with moderate $\mathcal{C}_0$ but weak $\mathcal{E}_1$ and $\mathcal{E}_2$ quickly plateau as structural constraints become stricter. For example, the strongest baseline, AG2, experiences a substantial degradation: its row and column accuracy drop to less than 50\% of its cell accuracy, and its table accuracy falls to below 5\% of its cell accuracy.
In contrast, \sysname maintains consistently high performance from cell to table to task accuracy, indicating that the three \paradigmname dimensions compose effectively rather than degrading at higher structural levels. 

From the ranking perspective, the rankings of agents at cell and task accuracy levels are exactly the same. This one-to-one correspondence indicates that $\mathcal{C}0$ serves as the backbone of end-to-end task success in \paradigmname tasks, with improvements in cell-level accuracy translating directly into gains in overall task performance.
 
Stratification emerges at the row, column, and table accuracy levels, corresponding to $\mathcal{E}1$ and $\mathcal{E}2$. At these higher structural levels, \sysname is exceptionally dominant, substantially outperforming all baselines. AG2, AutoGen, and AutoGPT form a second tier of sophisticated agentic frameworks with explicit multi-step planning and tool-use capabilities, but their performance drops sharply when moving from cell-level correctness to row- and table-level consistency. OpenAI ResearchBot operates at a noticeably lower tier as a comparatively lightweight research-oriented agent, with more constrained planning and reasoning mechanisms. As a result, it struggles to maintain coherent multi-cell reasoning beyond basic retrieval. WebVoyager and Open Deep Research remain at the bottom tier; these systems are more specialized for browsing or exploratory research scenarios rather than structured data construction, and they exhibit almost no reliable multi-cell or table-level reasoning capability.
This tiered pattern reveals an important insight: while many agents can occasionally retrieve correct individual values, none of them can systematically exploit structural dependencies across rows and columns or maintain global consistency across an entire table. 

Taken together, these findings validate the design of our evaluation framework: by combining strict reproducibility, semantic robustness, and structural stratification, our evaluation metrics provide reliable, discriminative, and interpretable comparisons across agents on \paradigmname tasks.

% Overall, these results demonstrate that existing web agents largely struggle with structured multi-level reasoning required by \paradigmname tasks, whereas \sysname successfully integrates in-depth information discovery ($\mathcal{C}0$), structural dependency modeling ($\mathcal{E}1$), and global table consistency ($\mathcal{E}2$) into a unified and scalable system.

\begin{figure}[t]
    \graphicspath{{figures/}}
    \centering
    \includegraphics[width=0.9\columnwidth]{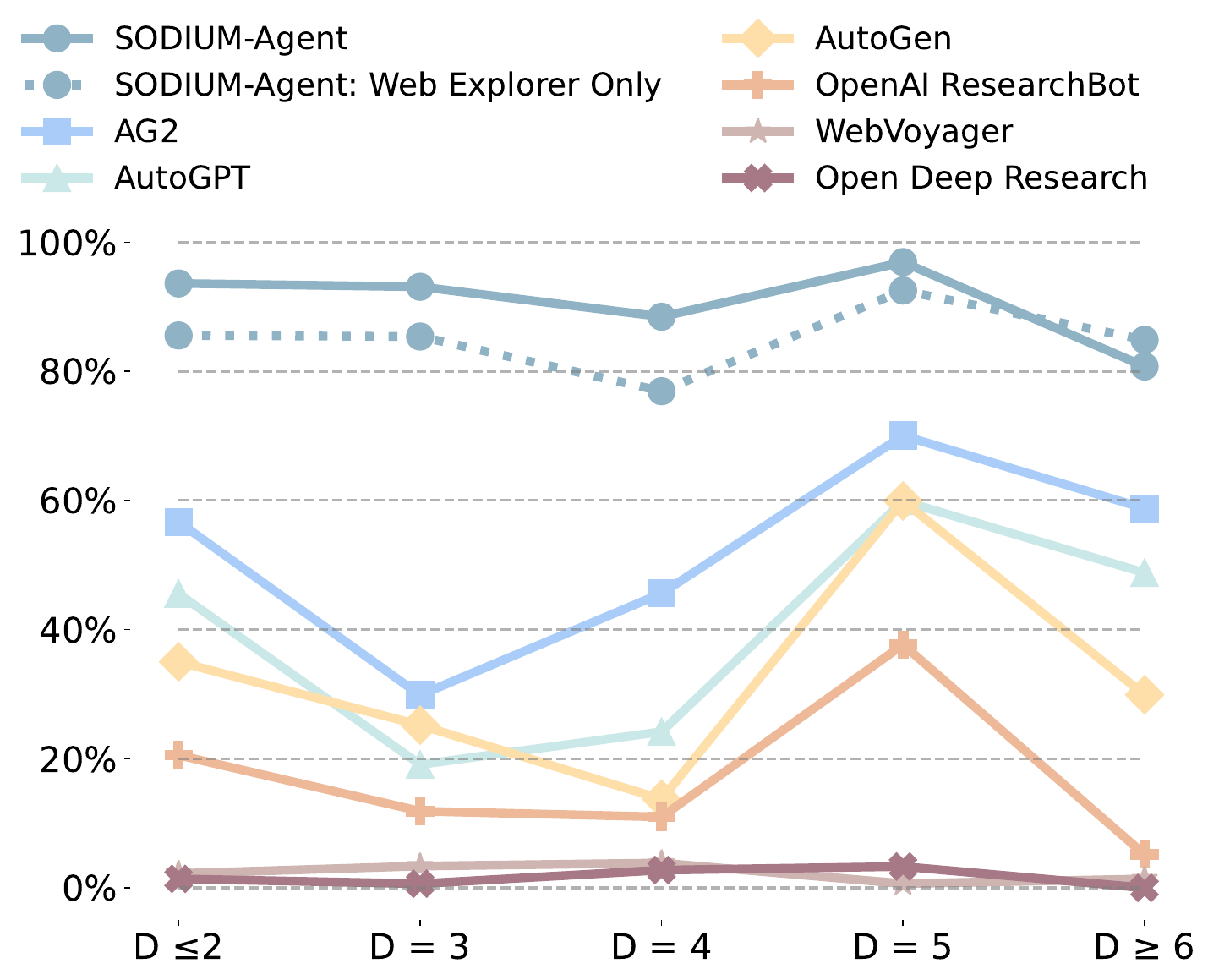}
    \caption{LLM-as-a-judge accuracy of different agents across different search depths $D$.}
    \label{fig:acc_by_depth}%
\end{figure}

\minihead{\sysname achieves consistently superior performance across search depths}
As we show in Figure~\ref{fig:acc_by_depth}, \sysname maintains stable and consistently high accuracy across all search depths, substantially outperforming all baselines.

Interestingly, performance does not monotonically decrease with depth for all agents. Instead, we observe a non-linear pattern: accuracy drops from shallow pages ($D \le 2$) to intermediate layers ($D=3$), then improves and peaks around $D=5$, before decreasing again at very deep levels ($D \ge 6$). This behavior reflects structural properties of real-world websites: agent performance drops from $D \le 2$ to $D=3$ due to increased exploration difficulty, as deeper navigation requires additional decisions, each introducing further opportunities for error. As search depth increases to $D=3$--$5$, navigation difficulty remains relatively stable. However, deeper pages tend to be more specialized and narrowly scoped. This reduces cross-topic interference and makes extraction more reliable once the correct page is reached, resulting in an increase in agent performance. However, at deeper levels ($D \ge 6$), exploration complexity becomes dominant again, and this long-horizon brittleness leads to the final decline in performance.

\subsection{\sysname's web explorer is efficient and robust}
\label{subsec:exp_web_explorer}
\begin{table}[t]
\centering
\small
\caption{The execution statistics of \sysname's web explorer across \algname levels $\ell$ with (as an embedded component of \sysname) and without the cache manager (as a standalone agent).}
\resizebox{0.9\columnwidth}{!}{
\begin{tabular}{llcccc}
\toprule
 & & $\ell=0$ & $\ell=1$ & $\ell=2$ & $\ell\ge 3$ \\
\midrule

\multirow{2}{*}{\# Cells}
& Embedded & 306 & 185 & 69 & 44 \\
& Standalone & 276 & 1006 & 524 & 343 \\

\midrule

\multirow{2}{*}{Accuracy (\%)}
& Embedded & 91.2 & 91.9 & 92.8 & 93.2 \\
& Standalone & 81.2 & 86.2 & 93.9 & 81.3 \\

\midrule

\multirow{2}{*}{Prune rate (\%)}
& Embedded & -- & 63.0 & 96.6 & 93.5 \\
& Standalone & -- & 78.8 & 96.9 & 93.6 \\

\midrule

\multirow{2}{*}{New URL rate (\%)}
& Embedded & -- & 35.8 & 29.3 & 25.7 \\
& Standalone & -- & 48.4 & 35.8 & 30.8 \\

\bottomrule
\end{tabular}
}
\label{tab:web_explorer_depth_filter}
\end{table}
As we show in Table~\ref{tab:web_explorer_depth_filter}, \sysname's web explorer achieves consistently high accuracy across \algname traversal levels, with most cells resolved at shallow levels.
Here, we use $\ell$ to refer to the traversal level to differentiate the search depth $D$ in Section \ref{subsec:overall_res}, with the latter referring to the intrinsic structural depth of the target cell values.

\minihead{\sysname resolves most cells with shallow exploration}
A large majority of cells are completed with $\ell\le 1$ (306 at $\ell=0$ and 185 at $\ell=1$), while only 113 cells require deeper exploration ($\ell\ge 2$).
This indicates that \sysname is generally able to identify the correct region of a website and navigate to the target sources quickly, either through cache-assisted inference ($\ell=0$) or with a single sorting iteration ($\ell=1$).
Combining the results in Table \ref{tab:web_explorer_depth_filter} and Table \ref{tab:cache_comparison}, we can see that out of the 500 cells that start from cached paths, in 306 of them (over 60\%), the cache manager directly outputs the correct source as the initial point of entry for the web explorer.
Across all cells, the average number of pages explored is 8.2 (min 1, max 172, median 3), suggesting that effective retrieval typically requires only a small number of page inspections.

\minihead{Accuracy is stable across search depths and page exploration}
Accuracy remains remarkably stable across depths, ranging from 91.2\% at $\ell=0$ to 93.2\% at $\ell\ge 3$.
Importantly, we observe no meaningful relationship between the number of pages visited and correctness: the Pearson correlation \cite{pearson1895regression} between page count and accuracy is $-0.063$, and the Spearman correlation \cite{spearman1904} is $0.054$, both well below 0.1.
This indicates that deeper or broader exploration does not inherently degrade accuracy; instead, correctness depends on structural reasoning rather than search length.

\minihead{Web explorer stabilizes the search space at deeper levels}
The prune rate (i.e., the proportion of candidate webpages discarded at each depth step) rises sharply from $\ell=1$ to $\ell=2$, coinciding with an approximately $K=10\times$ increase in the number of visited webpages. 
Beyond $\ell=2$, both the number of visited pages per level and the prune rate stabilize, remaining consistently high. This indicates that \algname effectively compresses the expanded search space and prevents uncontrolled growth as exploration proceeds, maintaining stable and selective traversal even at deeper levels.

Meanwhile, the new URL rate gradually decreases with depth (35.8\% $\rightarrow$ 29.3\% $\rightarrow$ 25.7\%). This suggests that URL synthesis is most beneficial in early exploration stages, when structural patterns are still being inferred. As depth increases and navigation becomes more grounded and stabilized in observed links, the need for speculative URL generation naturally diminishes.
% This reflects an increasingly aggressive pruning behavior.
% At early depths, the candidate space is broad and contains multiple plausible directions, so the system retains a substantial portion of candidates.
% However, once structural signals become clearer, the explorer confidently eliminates most remaining candidates, resulting in very high pruning rates at deeper levels.

\minihead{Dynamic webpages dominate web exploration}
Across all cells and stages, \texttt{inspect\_dynamic} constitutes the vast majority (91.18\%) of page inspections.
\sysname also typically \emph{finishes} on dynamic pages: the last inspection call is dynamic for 1{,}901 cells versus 220 static and 28 online documents. Accuracy is slightly higher when the final inspection is dynamic (92.43\%) than when it is static (90.91\%), suggesting that dynamic endpoint pages are reliable for extracting the target values.

\subsection{\sysname's cache manager improves accuracy and reduces costs}
\label{subsec:exp_cache}
\begin{table}[t]
    \caption{
Comparison of \sysname components in the utility of cached results during cell filling. Accuracy is measured using LLM-as-a-judge evaluation, which provides finer-grained assessment, as analyzed in Section~\ref{subsec:overall_res}.
}
\centering
\small
\resizebox{\linewidth}{!}{
\begin{tabular}{lc|c|c|c}
\toprule
\textbf{Component}
& \textbf{Mode}
& \textbf{\# Cells}
& \textbf{Acc. (\%) $\uparrow$} 
& \textbf{Cost (\$) $\downarrow$} \\
\midrule
\multirow{2}{*}{\shortstack{Web\\Explorer}}
    & Start from base           & 104 & 87.50 & 1.08 \\
    & Start from cached paths   & 500 & 92.60 & 0.91 \\
\midrule
\multirow{2}{*}{\shortstack{Cache\\Manager}}
    & Left    & 885 & 93.67 & 0.23 \\
    & Up      & 660 & 89.24 & 0.16 \\
\bottomrule
\end{tabular}
}
\label{tab:cache_comparison}
\end{table}

We report the cache utility results in Table~\ref{tab:cache_comparison}. 
Across the total of 2,149 evaluated cells, we have the following observations.

\minihead{Cached results are heavily used}
Out of 2,149 cells, 1,545 cells (885 Left + 660 Up) are answered directly by the Cache Manager, accounting for 71.9\% of all cells.
An additional 500 cells (23.27\%) use cached navigation paths before invoking the web explorer.
Only 104 cells (4.8\%) start exploration directly from the base URL without any cache assistance.
Overall, 2045 out of 2,149 cells (95.2\%) benefit from caching, demonstrating that cache reuse is the dominant execution pattern in \sysname.

\minihead{Cache improves accuracy}
Direct cache usage (Left + Up) achieves the highest accuracy.
The Left cache direction reaches 93.7\%, while the Up direction achieves 89.2\%.
Both outperform starting from base (87.5\%).
Even when the system must fall back to web exploration, starting from cached paths improves accuracy from 87.5\% to 92.6\%, a gain of 5.1 percentage points.
These results demonstrate that structural reuse strengthens both efficiency and robustness: the cache manager reduces redundant traversal while simultaneously increasing extraction accuracy.

\minihead{Cache reduces cost}
To quantify efficiency, we measure the API cost incurred per cell. 
Direct cache retrieval substantially lowers exploration cost.
On average, the Left and Up cache directions incur only \$0.25 and \$0.18 per cell, respectively.
In contrast, web exploration starting from the base page costs \$1.08 per cell, which is 19\% higher than exploration initialized from cached paths.
These results demonstrate that structural reuse effectively reduces browsing overhead and overall token expenditure.

\minihead{Cached path search effectively balances reuse and synthesis}
Across cells that use cached paths, \sysname proposes an average of 4.44 candidate URLs per call. Notably, 60.4\% of these URLs are newly synthesized rather than directly observed from the current page context. This indicates that the cache manager does not merely reuse previously discovered links, but actively generalizes structural patterns to generate plausible navigation paths. 
The relatively small proposal set keeps the branching factor low, while the high proportion of synthesized URLs allows \sysname to actively adapt to unseen yet structurally similar cases.

\vspace{1em}

Overall, these results show that the cache manager is responsible for handling nearly two-thirds of all cells directly, improving both accuracy and efficiency, while cached path reuse further enhances robustness when exploration is unavoidable.

\subsection{$\mathcal{E}1$ and $\mathcal{E}2$ are critical for \sysname}
\label{subsec:exp_ablation}
We ablate the cache manager and run the web explorer in isolation to assess its standalone performance and quantify the incremental contribution of caching. The standalone web explorer already outperforms all baseline agents by a substantial margin, and the cache manager consistently lowers the cost of \sysname while further improving accuracy, reinforcing the conclusions drawn in Section~\ref{subsec:exp_cache}. We detail our findings below.

\minihead{The standalone web explorer substantially outperforms all baselines}
As we show in Table~\ref{tab:overall_res}, the web explorer alone significantly outperforms all baseline agents across every metric. In particular, it achieves 84.37\% task-level accuracy under LLM-as-a-judge evaluation, compared to 46.48\% for the strongest baseline, AG2, corresponding to a $1.82\times$ improvement.

This advantage is consistent across structural levels. At the cell level, the standalone web explorer achieves over $1.9\times$ the accuracy of AG2. At the row and column levels, the gains are even more pronounced: over $4\times$ at the row level and over $3\times$ at the column level. At the table level, where multi-cell consistency is critical, the web explorer achieves a $17\times$ improvement over AG2.

As the web explorer constitutes the core retrieval component of \sysname, these results reinforce our earlier claim that $\mathcal{C}_0$ serves as a foundational capability. Once reliable deep retrieval is established, substantial performance gains over existing agentic systems can already be realized.

\minihead{Cache manager and web explorer synergize to achieve strong structural integrity}
Integrating the cache manager further improves performance across all levels. 
Task accuracy increases from 84.37\% to 91.07\%, a gain of 8\% relative improvement. 

The gains become more pronounced at stricter structural levels. 
At the cell level, accuracy increases by 5.12 percentage points, while row accuracy increases by 14.68 percentage points, column accuracy by 14.05 percentage points, and table accuracy by 20.95 percentage points.
This demonstrates that while $\mathcal{C}0$ enables strong cell-wise retrieval, $\mathcal{E}1$ and $\mathcal{E}2$, which leverage structural dependencies and reuse discovered paths, are critical for maintaining global coherence across rows and entire tables.

In short, the web explorer provides depth and coverage, while the cache manager provides structural stability and consistency. 
Their combination allows \sysname to exceed 90\% task-level accuracy while dramatically improving table-level correctness.

\minihead{Cache manager stablizes \sysname's performance across different search depths}
As we illustrate in Figure \ref{fig:acc_by_depth}, the standalone web explorer exhibits noticeably larger performance fluctuations than \sysname. The cache manager reduces this instability by reusing validated paths and previously discovered source formats. As a result, \sysname consistently outperforms the standalone variant across most depths, with only minor differences at very large depths. This is as expected because as depth increases, the search space becomes more fragmented, with a higher branching factor; consequently, path reuse is more susceptible to minor structural divergence. However, this marginal degradation at extreme depths is insignificant relative to the pronounced overall gains delivered by structural reuse.
% As depth increases and webpage structures become more repetitive, caching enables effective reuse of recurring templates instead of re-exploration, leading to smoother and more stable performance overall.

\minihead{Cache manager significantly improves efficiency}
Beyond accuracy, the cache manager substantially reduces exploration overhead. 
As we show in Table~\ref{tab:web_explorer_depth_filter}, the \algname traversal level distribution shifts markedly when the cache manager is removed.

In the embedded setting, 81.3\% of cells are resolved within traversal levels 0 and 1, with only 18.7\% requiring traversal level 2 or beyond. 
In contrast, without caching, the proportion of deeper explorations ($\ell\ge2$) rises to 40.4\%, more than doubling.
The prune rates at all levels also increase in the standalone setting, indicating a larger fraction of discarded intermediate states and thus more ineffective browsing steps. 
Moreover, the new URL rate is consistently higher at all levels without caching, suggesting that the agent is more frequently navigating to previously unseen pages rather than reusing established structural patterns. 
This behavior reflects a lack of access to cached source formats and leads to more exploratory and redundant navigation.

This shift toward deeper exploration directly translates to higher cost. 
The average cost per cell is \$1.36 for the standalone web explorer and \$0.41 for \sysname. 
Introducing the cache manager reduces cost by approximately 70\%.

% The task-level accuracy of \sysname drops from 91.1\% to 84.3\%. Although the system still outperforms all baselines, this represents a substantial performance degradation, demonstrating the critical role of the cache manager in maintaining high task-level consistency and efficiency.

% We further replace the underlying model with the newly released Claude Opus 4.6. Specifically, we randomly select one task from each domain and evaluate performance compared to GPT-5. We observe no significant oscillations across domains, suggesting that the performance trends are stable and not strongly tied to a specific frontier model.
% 2, 15, 20, 29, 99, 106
\section{Related Work}
\label{sec:related_work}

We review related work from the following three aspects.

\minihead{LLM for Data Management}
Data science workflows have traditionally been labor-intensive and iterative \cite{8440815}. 
With the rapid development of LLMs, recent work has explored augmenting different stages of the data lifecycle using LLM agents and models. These efforts range from correcting and imputing missing values in tables \cite{lakefill}, transforming unstructured data into structured tables \cite{Hu_2024, patel2025semanticoperatorsdeclarativemodel, shankar2025docetlagenticqueryrewriting, wei2026multiobjectiveagenticrewritesunstructured, biswal2024text2sqlenoughunifyingai}, aggregating tables from multiple sources \cite{Freire2025LargeLM}, to enabling downstream analytical workflows via NL2SQL \cite{lei2024spider, li2024can, shkapenyuk2025automaticmetadataextractiontexttosql} or more complex data science and machine learning pipelines \cite{hu2025reprobenchagenticaisystems, nam2025dsstardatascienceagent, nam2025mlestarmachinelearningengineering, zhang2025coddllmempoweringlargelanguage}.

Despite this progress, large-scale data collection from the open web, often the critical first step in real-world data science pipelines \cite{data-science-life-cycle, Hazzan2023}, remains relatively underexplored in terms of automation. Existing approaches, whether based on traditional web scraping techniques \cite{dsdd} or LLM agents \cite{Hu_2025}, typically assume that the target dataset is already well-organized and structurally complete on the web. In practice, however, relevant information is often scattered across heterogeneous webpages and requires integration and normalization, similar to constructing structured repositories such as data lakes. Addressing this gap is precisely the focus of \paradigmname.

\minihead{Open domain search tools}
Search tools substantially enhance the performance of LLM agents by providing access to open-domain knowledge \cite{autogpt_repo,ag2_docs,opendeepresearch,google_deep_research_2024,microsoft_autogen}. 
Modern web search tools integrated with LLMs also demonstrate promising capabilities \cite{claude_web_search_docs, gemini_web_search_docs, openai_web_search_docs}. 

Nevertheless, existing approaches exhibit two critical limitations when deep, domain-specific information is required. 
First, they are primarily optimized for general-purpose search. For example, the search engine \cite{fastapi_docs} used to improve SearchR1 \cite{jin2025searchr1trainingllmsreason} targets broad, generic queries rather than structured, domain-specific data extraction tasks. 
Second, they lack sufficient exploration depth. WebVoyager \cite{he2024webvoyager} performs relatively shallow website navigation, and other search-based systems largely rely on external search engines without explicitly optimizing multi-step, in-site exploration strategies. Existing web scrapers are effective for static websites \cite{jina_reader_api,scrapy,go_colly,crawlee}; however, real-world websites often require interactive behaviors, such as expanding tabs or triggering dynamic content loading, to fully access relevant information.
As a result, these approaches are insufficient for \paradigmname tasks, where answering a single query requires deep, structured traversal within specialized websites.

\minihead{Information retrieval and RAG systems.}
Information retrieval and RAG systems have demonstrated strong effectiveness across diverse data modalities, including structured data \cite{fang-etal-2025-tabgen}, unstructured text \cite{gao2024retrievalaugmentedgenerationlargelanguage, xiong2025raggymsystematicoptimizationlanguage, lewis2021retrievalaugmentedgenerationknowledgeintensivenlp}, and graph data \cite{NEURIPS2024_efaf1c97}. 
However, these approaches typically decouple retrieval from the structural organization of the target outputs, treating retrieval as an independent pre-processing step. 
Such separation is suboptimal for \paradigmname tasks, where retrieval decisions and structural reasoning must be tightly integrated.

Moreover, even methods that study dynamic or multi-frame data settings \cite{Chen_2023} assume that all relevant data has already been pre-extracted and stored locally. 
That is, they operate over a static, predefined corpus and do not require active, multi-step retrieval from live webpages. 

\section{Conclusion}
\label{sec:conclusion}
In this paper, we formalize the problem of structuring open-domain unstructured data into materialized databases as the \paradigmname task.
To quantitatively evaluate \paradigmname in real-world settings, we collect \datasetname of 105 queries from 6 domains.
To address this challenge, we develop \sysname, an agentic system composed of a web explorer and a cache manager. We design a novel \algname algorithm for \paradigmname tasks as a workflow for the web explorer to achieve deep, schema-driven exploration. The cache manager uses structural regularities across table cells to reuse validated navigation paths, enforce cross-cell consistency, and improve efficiency.
We evaluate \sysname on \datasetname and show that it achieves 91.1\% task-level accuracy, significantly outperforming state-of-the-art baselines. 
These results demonstrate that treating open domains as latent, materializable databases is both feasible and essential for supporting scalable, real-world analytical workflows.
%\clearpage

\bibliographystyle{ACM-Reference-Format}
\bibliography{sample}

\end{document}